\documentclass[a4paper,11pt]{article}
\RequirePackage{times}
\RequirePackage{courier}
\RequirePackage{mathptm}
\RequirePackage{graphicx}
\begin{document}

\title{Finite automata models of quantized systems:\\conceptual status and outlook}
\author{Karl Svozil\\
 {\small Institut f\"ur Theoretische Physik, University of Technology Vienna }     \\
  {\small Wiedner Hauptstra\ss e 8-10/136,}
  {\small A-1040 Vienna, Austria   }            \\
  {\small e-mail: svozil@tuwien.ac.at}}

\maketitle
\begin{abstract}
Since Edward Moore, finite automata theory has been inspired
by physics, in particular by quantum complementarity. We review
automaton complementarity, reversible automata and the connections
to generalized urn models. Recent developments in quantum information theory
may have appropriate formalizations in the automaton context.
\end{abstract}

\section{Physical connections}

Physics and computer science share common interests.
They may pursue their investigations by different methods and formalisms,
but once in a while it is quite obvious that the interrelations are pertinent.
Take, for example, the concepts of information and computation.
Per definition, any theory of information and computation,
in order to be applicable, should refer to
physically operationalizable concepts.
After all, information and computation is physical \cite{landauer}.

Conversely, concepts of computer science have increasingly
influenced physics. Two examples for these developments have been
the recent developments in classical continuum theory, well known
under the term ``deterministic chaos,''
as well as quantum information and computation theory.
Quantum systems nowadays are often perceived as very specific and delicate
(due to decoherence; i.e., the irreversible loss of state information in measurements)
reversible computations.

Whether or not this correspondence resides in the very foundations of both sciences
remains speculative. Nevertheless, one could conjecture a correspondence principle by
stating that
{\em every feature of a computational model should be reflected by some physical system.
Conversely, every physical feature, in particular of a physical theory,
should correspond to a feature of an appropriate
computational model.}
This is by no means trivial, as for instance the abundant use of nonconstructive continua
in physics indicates.
No finitely bounded computation could even in principle store, process
and retrieve the nonrecursively enumerable
and even algorithmically incompressible random reals,
of which the continuum ``mostly'' exists.
But also recent attempts to utilize quantum computations for speedups or even to solve
problems which are unsolvable within classical recursion theory \cite{2002-cal-pav}
emphasize the interplay between
physics and computer science.

Already quite early, Edward Moore attempted a formalization of quantum complementarity
in terms of finite deterministic automata \cite{e-f-moore}.
Quantum complementarity is the feature of certain microphysical systems
not to allow the determination of all of its properties with arbitrary precision at once.
Moore was interested in the initial state determination problem: given
a particular finite automaton which is in an unknown initial state; find that
initial state by the analysis of input-output experiments on a single such automaton.
Complementarity manifests itself if different inputs yield different properties
of the initial automaton state while at the same time steering the automaton into
a state which is independent of its initial one.

Moore's considerations have been extended in many ways.
Recently, different complementarity classes have been characterized
\cite{cal-sv-yu} and their likelihood  has been
investigated \cite{e-calude-lip,cal-cal-k}.
We shall briefly review
a calculus of propositions referring to the initial state problem
which resembles quantum logic in many ways \cite{svozil-93,svozil-ql}.
Automaton theory can be liked to generalized urn models
\cite{svozil-2001-eua}.
In developing the analogy to quantum mechanics further,
reversible deterministic finite automata have been introduced
\cite{sv-aut-rev}.
New concepts in quantum mechanics \cite{zeil-99} suggest
yet different finite automaton models.

\section{Automaton partition logics}
Consider a Mealy automaton
$\langle S,I,O,\delta ,\lambda \rangle$, where
 $S,I,O$ are the sets of states,
input and output symbols, respectively.
$\delta (s,i)=s'$ and
$\lambda (s,i)=o$,
$s,s'\in S$,
$i\in I$
and $o\in O$
are the transition and the output functions, respectively.

The initial state determination problem can be formalized as follows.
Consider a particular automaton
and all sequences of input/output symbols which result from
all conceivable experiments on it.
These experiments induce a state partition in the following natural way.
Every distinct set of input/output symbols is associated with a set of initial automaton
states which would reproduce that sequence.
This set of states may contain one or more states, depending on the ability of the experiment
to separate different initial automaton states.
A partitioning of the automaton states associated with an input sequence
is obtained if one considers the variety of all possible output sequences.
Stated differently:
given a set of input symbols, the set of automaton states ``decays'' into disjoint
subsets associated with the possible output sequences.
This partition can then be identified with a Boolean algebra,
with the elements of the partition interpreted as atoms.
By pasting the Boolean algebras of the ``finest'' partitions together, one obtains
a calculus of proposition associated with the particular automaton.
This calculus of propositions is referred to as {\em automaton partition logic.}

The converse is true as well:
given any partition logic, it is always possible to (nonuniquely)
construct a corresponding automaton with the following specifications:
associate with every element of the set of partitions a single input symbol.
Then take the partition with the highest number of elements and associate a single output
symbol with any one element of this partition.
The automaton output function  can then be defined by associating a single output symbol per element
of the partition (corresponding to a particular input symbol).
Finally, choose a transition function which completely looses the state information
after only one transition; i.e., a transition function which maps all automaton state into
a single one.
We just mention that another, independent, way to obtain automata from partition logics
is by considering the set of two-valued states.

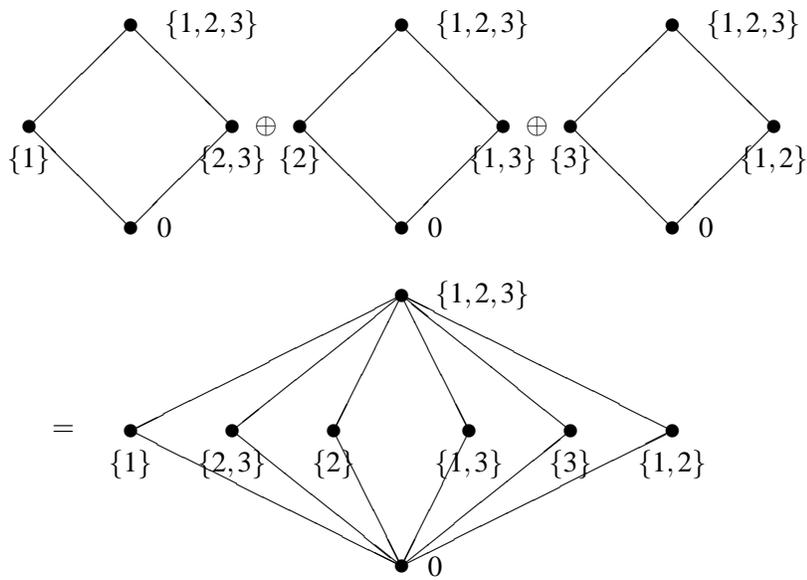
\begin{figure}
\begin{center}
\unitlength 0.90mm
\linethickness{0.4pt}
\begin{picture}(110.00,91.00)
\put(0.00,75.00){{\circle*{2.00}}                                  }
\put(15.00,60.00){{\circle*{2.00}}                                 }
\put(15.00,90.00){{\circle*{2.00}}                                 }
\put(30.00,75.00){{\circle*{2.00}}                                 }
\put(0.00,75.00){{\line(1,1){15.00}}                               }
\put(15.00,90.00){{\line(1,-1){15.00}}                             }
\put(30.00,75.00){{\line(-1,-1){15.00}}                            }
\put(15.00,60.00){{\line(-1,1){15.00}}                             }
\put(35.00,75.00){\makebox(0,0)[cc]{$\oplus$}}
\put(20.00,90.00){{\makebox(0,0)[lc]{$\{1,2,3\}$}}                 }
\put(20.00,60.00){{\makebox(0,0)[cc]{$0$}}                 }
\put(30.00,70.00){{\makebox(0,0)[cc]{$\{2,3\}$}}                   }
\put(0.00,70.00){{\makebox(0,0)[cc]{$\{1\}$}}                      }
\put(40.00,75.00){{\circle*{2.00}}                               }
\put(80.00,75.00){{\circle*{2.00}}                        }
\put(55.00,60.00){{\circle*{2.00}}                               }
\put(95.00,60.00){{\circle*{2.00}}                        }
\put(55.00,90.00){{\circle*{2.00}}                               }
\put(95.00,90.00){{\circle*{2.00}}                        }
\put(70.00,75.00){{\circle*{2.00}}                               }
\put(110.00,75.00){{\circle*{2.00}}                       }
\put(40.00,75.00){{\line(1,1){15.00}}                            }
\put(80.00,75.00){{\line(1,1){15.00}}                     }
\put(55.00,90.00){{\line(1,-1){15.00}}                           }
\put(95.00,90.00){{\line(1,-1){15.00}}                    }
\put(70.00,75.00){{\line(-1,-1){15.00}}                          }
\put(110.00,75.00){{\line(-1,-1){15.00}}                  }
\put(55.00,60.00){{\line(-1,1){15.00}}                           }
\put(95.00,60.00){{\line(-1,1){15.00}}                    }
\put(75.00,75.00){\makebox(0,0)[cc]{$\oplus$}}
\put(60.00,90.00){{\makebox(0,0)[lc]{$\{1,2,3\}$}}               }
\put(100.00,90.00){{\makebox(0,0)[lc]{$\{1,2,3\}$}}       }
\put(60.00,60.00){{\makebox(0,0)[cc]{$0$}}               }
\put(100.00,60.00){{\makebox(0,0)[cc]{$0$}}       }
\put(70.00,70.00){{\makebox(0,0)[cc]{$\{1,3\}$}}                 }
\put(110.00,70.00){{\makebox(0,0)[cc]{$\{1,2\}$}}         }
\put(40.00,70.00){{\makebox(0,0)[cc]{$\{2\}$}}                   }
\put(80.00,70.00){{\makebox(0,0)[cc]{$\{3\}$}}            }
\put(55.00,50.00){\circle*{2.00}}
\put(55.00,10.00){\circle*{2.00}}
\put(55.00,50.00){{\line(-2,-1){40.00}}                            }
\put(15.00,30.00){{\line(2,-1){40.00}}                             }
\put(55.00,10.00){{\line(2,1){40.00}}                     }
\put(95.00,30.00){{\line(-2,1){40.00}}                    }
\put(55.00,50.00){{\line(-5,-4){25.00}}                            }
\put(30.00,30.00){{\line(5,-4){25.00}}                             }
\put(55.00,10.00){{\line(5,4){25.00}}                     }
\put(80.00,30.00){{\line(-5,4){25.00}}                    }
\put(55.00,50.00){{\line(-1,-2){10.00}}                          }
\put(45.00,30.00){{\line(1,-2){10.00}}                           }
\put(55.00,50.00){{\line(1,-2){10.00}}                           }
\put(65.00,30.00){{\line(-1,-2){10.00}}                          }
\put(15.00,30.00){{\circle*{2.00}}}
\put(30.00,30.00){{\circle*{2.00}}}
\put(45.00,30.00){{\circle*{2.00}}                               }
\put(65.00,30.00){{\circle*{2.00}}                               }
\put(80.00,30.00){{\circle*{2.00}}                        }
\put(95.00,30.00){{\circle*{2.00}}                        }
\put(30.00,25.00){\makebox(0,0)[cc]{{$\{2,3\}$}}}
\put(45.00,25.00){\makebox(0,0)[cc]{{$\{2\}$}}                   }
\put(65.00,25.00){\makebox(0,0)[cc]{{$\{1,3\}$}}                 }
\put(80.00,25.00){\makebox(0,0)[cc]{{$\{3\}$}}            }
\put(95.00,25.00){\makebox(0,0)[cc]{{$\{1,2\}$}}          }
\put(15.00,25.00){\makebox(0,0)[cc]{{$\{1\}$}}   }
\put(60.00,10.00){\makebox(0,0)[cc]{$0$}}
\put(60.00,50.00){\makebox(0,0)[lc]{$\{1,2,3\}$}}
\put(5.00,30.00){\makebox(0,0)[cc]{$=$}}
\end{picture}
\end{center}
\caption{ \label{xx1} Automaton partition logic
corresponding to the Mealy automaton
$\langle \{1,2,3\},\{1,2,3\},\{0,1\},\delta = 1 ,\lambda(s,i)=\delta_{si} \rangle$.}\end{figure}

In that way, a multitude of worlds can be constructed,
many of which feature quantum complementarity.
For example, consider the Mealy automaton
$\langle \{1,2,3\},\{1,2,3\},\{0,1\},\delta = 1 ,\lambda(s,i)=\delta_{si} \rangle$
(the Kronecker function $\delta_{si}=1$ if $s=i$, and zero otherwise).
Its states
are partitioned into
$
\{\{1\},\{2,3\}\}
$,
$
\{\{2\},\{1,3\}\}
$,
$
\{\{3\},\{1,2\}\}
$, for the inputs 1, 2, and 3, respectively.
Every partition forms a Boolean algebra $2^2$.
The partition logic depicted in Fig. \ref{xx1}
is obtained by ``pasting'' the three algebras together; i.e., by maintaining the order structure and by
identifying
identical elements; in this case $\emptyset,\{3,1,2\}$.
It is a modular, nonboolean lattice  ${\rm MO}_3$ of the ``chinese lantern'' form.
A systematic study \cite[pp.~38-39]{svozil-ql}
shows that automata reproduce (but are not limited to)
all finite subalgebras of Hilbert lattices of finite-dimensional quantum logic.

Mealy automata are logically equivalent
to generalized urn model (GUM) \cite{wright:pent,wright}
$\langle U,C,L,\Lambda \rangle $
which is an ensemble $U$ of ball types with black background color.
Printed on these balls are some symbols from a symbolic alphabet $L$.
These symbols are colored.
The colors are elements of a set of colors $C$.
A particular ball type is associated with a unique combination of mono-spectrally
(no mixture of wavelength) colored symbols
printed on the black ball background.

Suppose you have a number of colored eyeglasses built from filters for the
$\vert C\vert $ different colors.
They should  absorb every other light than one of a particular single color.
When a spectator looks at a particular ball through such an eyeglass,
the only recognizable symbol will be the one in the particular
color which is transmitted through the eyeglass.
All other colors are absorbed, and the symbols printed in them will appear black
and therefore cannot be differentiated from the black background.
Hence the ball appears to carry a different ``message'' or symbol,
depending on the color at which it is viewed.
The above procedure could be formalized by a  ``lookup'' function $\Lambda (u,c)=v$,
which depends on the ball type $u\in U$ and on the color
$c\in C$, and which returns the symbol
$v\in L$ printed in this color.
(Note the analogy to the output function $\lambda$.)

Again, an empirical logic can be constructed as follows.
Consider the set of all ball types.
With respect to a particular colored eyeglass, this set
gets partitioned into those ball types which can be separated by the particular color of
the eyeglass.
Every such state partition can then be identified with a Boolean algebra whose atoms are the elements of the partition.
A pasting of all of these Boolean algebras yields the calculus of propositions.
The corresponding correlation polytope formed by the set of two-valued states
characterizes all probability measures.

In order to define an automaton partition logic associated with a Mealy automaton
$\langle S,I,O,\delta ,\lambda \rangle$
from a GUM $\langle U,C,L,\Lambda \rangle $,
let
$u\in U$,
$c\in C$,
$v\in L$,
and
$s,s'\in S$,
$i\in I$,
$o\in O$, and assume
$\vert U\vert =\vert S\vert$,
$\vert C\vert =\vert I\vert$,
$\vert L\vert =\vert O\vert$.
The following identifications can be made
with the help of  the bijections $t_S,t_I$ and $t_O$:
\begin{equation}
\begin{array}{llllll}
t_S(u)=s, \;
 t_I(c)=i, \;
 t_O(v)=o, \\
\delta (s,i)= s_i  \quad {\rm for \; fixed\;}s_i\in S {\rm \;and \;
arbitrary\;}s\in S,\; i\in I,\\
\lambda  (s,i) = t_O\left(\Lambda (t_S^{-1}(s),t_I^{-1}(i))\right).
\end{array}
\label{oto-ua}
\end{equation}

Conversely,
consider an arbitrary Mealy automaton $\langle S,I,O,\delta ,\lambda \rangle$.
Just as before, associate with every single automaton state $s\in S$
a ball type $u$,
associate with every input symbol $i\in I$
a unique color $c$,
and
associate with every output symbol $o\in O$
a unique symbol $v$; i.e., again
$\vert U\vert =\vert S\vert$,
$\vert C\vert =\vert I\vert$,
$\vert L\vert =\vert O\vert$.
The following identifications can be made
with the help of  the bijections $\tau_U,\tau_C$ and $\tau_L$:
\begin{eqnarray}
\begin{array}{llllll}
\tau_U(s)=u, \; \tau_C(i)=c, \;  \tau_L(o)=v, \;
\Lambda  (u,c) = \tau_L (\lambda (\tau_U^{-1}(u),\tau_C^{-1}(c))).
\end{array}
\label{oto-au}
\end{eqnarray}
A direct comparison of
(\ref{oto-ua})
and
(\ref{oto-au})
yields
\begin{eqnarray}
\begin{array}{llllll}
\tau_U^{-1}=t_S, \; \tau_C^{-1}=t_I, \;  \tau_L^{-1}=t_O.
\end{array}
\label{oto-oto}
\end{eqnarray}

For example, for the automaton partition logic
depicted in Fig. \ref{xx1}, the above  construction yields
a GUM with three ball types:
type 1 has a red ``0'' printed on it and a green and blue ``1;''
type 2 has a green ``0'' printed on it and a red and blue ``1;''
type 3 has a blue ``0'' printed on it and a red and green ``1.''
Hence, if an experimenter looks through a red eyeglass, it would be possible
to differentiate between
ball type 1 and the rest (i.e., ball types 2 and 3).

\section{Reversible finite automata}
The quantum time evolution between two measurements is reversible,
but the previously discussed Moore and Mealy automata are not.
This has been the motivation for another type of deterministic finite automaton
model \cite{sv-aut-rev,svozil-ql}
whose combined transition and output function is bijective and thus reversible.

The elements of the Cartesian product
$S\times I$ can be arranged as a linear list $\Psi$ of length
$n$, just like a vector.
Consider the automaton $\langle S,I,I,\delta ,\lambda \rangle \equiv \langle \Psi , U \rangle$
with
$I=O$ and $U:(s,i)\rightarrow (\delta(s,i),\lambda (s,i))$,
whose time evolution can in be rewritten as
\begin{equation}
\sum_{j=1}^n U_{ij}\Psi_j
\end{equation}
$U$ is a $n\times n$-matrix which, due to the requirement
of determinism, uniqueness and invertability is a {\em permutation matrix.}
The class of all reversible automata corresponds to the group of permutations,
and thus reversible automata are characterized by permutations.

Do reversible automata feature complementarity?
Due to reversibility, it seems to be always possible to measure
a certain aspect of the initial state problem,
copy this result to a ``save place,''
revert the evolution and measure an arbitrary other property.
In that way, one effectively is able to work with an arbitrary number of automaton copies
in the same initial state.
Indeed, as already Moore has pointed out, such a setup cannot yield complementarity.

The environment into which the  automaton is embedded is of conceptual importance.
If the environment allows for copying;
i.e., one-to-many evolutions, then the above argument applies and there is no complementarity.
But if this environment is reversible as well,
then in order to revert the automaton back to its original initial state,
all information has to be used which has been acquired so far; i.e., all changes
have to be reversed, both in the automaton as well as in the environment.
To put it pointedly: there is no  ``save place'' to which the result
of any measurement could be copied and stored and afterwards retrieved
while the rest of the system returns to its former state.
This is an automaton analogue to the no-cloning theorem of quantum information theory.

For example, a reversible automaton corresponding to the permutation whose cycle form
is given by (1,2)(3,4) corresponds to an automaton
$\langle \{1,2\},\{0,1\},\{0,1\},\delta (s,i)=s ,\lambda (s,i)= (i+1)\,{\rm mod} \,2 \rangle$,
or, equivalently,
\begin{equation}
\left\langle
\left(
\begin{array}{llllll}
(1,0)\\(1,1)\\(2,0)\\(2,1)
\end{array}
\right) ,
\left(
{
\begin{array}{llllll}
0&1&0&0\\
1&0&0&0\\
0&0&0&1\\
0&0&1&0
\end{array}
}
\right)
  \right\rangle .
\end{equation}
A pictorial representation may be given in terms of a flow diagram
which represent the permutation of the states in one evolution step.

\section{Counterfactual automata}

Automaton partition logic is nondistributive and thus nonclassical (if Boolean logic is considered
to be classical).
Yet it is not nonclassical as it should be, in particular as compared to quantum mechanics.
The partitions have a set theoretic interpretation,
and automaton partition logic allows for a full set of two-valued states;
i.e., there are a sufficient number of two-valued states for constructing
lattice homomorphisms.
This is not the case for Hilbert lattices associated with quantum mechanical
systems of dimension higher than two.
Probably the most striking explicit and constructive demonstration of this fact
has been given by Kochen and Specker \cite{kochen1}.
They enumerated a finite set of interconnected orthogonal tripods
in threedimensional real Hilbert space
with the property that the tripods cannot be colored
consistently in such a way that two axes are green and one is red.
Stated differently, the chromatic number of the threedimensional real unit sphere is four.
The entire generated system of associated properties is not finite.
Just as for all infinite models, denumerable or not,
it does not correspond to any finite automaton model.

Since the colors ``green'' and ``red'' can be interpreted as the truth values
``false'' and ``true'' of properties of specific quantum mechanical systems,
there has been much speculation as to the existence of complementary quantum physical properties.
One interpretation supposes that counterfactual properties
different from the ones being measured do not exist.
While this discussion may appear rather philosophical and not to the actual physical point,
it has stimulated the idea that a particle does not carry more (counterfactual)
information than it has been prepared for.
I.e., {\em a $n$-state particle carries exactly one nit of information,
and $k$ $n$-state particles carry exactly $k$ nits of information}
\cite{zeil-99,DonSvo01,svozil-2002-statepart-prl}.
As a consequence, the information may not be coded into single particles alone (one nit per particle),
but in general may be distributed over ensembles of particles, a property called entanglement.
Furthermore, when measuring properties different from the ones the particle is
prepared for, the result may be irreducible random.
The binary case is $n=2$. Note that, different from classical continuum theory,
where the base of the coding is a matter
of convention and convenience, the base of quantum information
is founded on the $n$-dimensionality of Hilbert space of the quantized system,
a property which is unique and measurable.

A conceivable automaton model is one which contains a set of input symbols $I$ which themselves
are sets with $n$ elements corresponding to the $n$ different outputs of the $n$-ary information.
Let
$\langle I\times O,I,O,\delta ,\lambda \rangle$
with $S=I$ to reflect the issue of state preparation versus measurement.
Since the information is in base $n$, $O=\{1,\ldots ,n\}$.
Furthermore, require that after every measurement, the automaton is in a state
which is characterized by the measurement; and that random results are obtained if
different properties are measured than the ones the automaton has been prepared in. I.e.,
\begin{equation}
\begin{array}{llllll}
\delta ((s,o_s),i)=\left\{
{
\begin{array}{llllll}
(i,{\rm RANDOM}(O)) \quad {\rm if}\; s\neq i,\\
 (s,o_s) \quad {\rm if}\; s = i,
\end{array}
}
\right. ,\\
\lambda ((s,o_s),i)=\left\{
{
\begin{array}{llllll}
{\rm RANDOM}(O) \quad {\rm if}\; s\neq i,\\
 o_s \quad {\rm if}\; s = i,
\end{array}
}
\right.
\end{array}
\end{equation}
The state $(s,o_s)\in I\times O$
is characterized by the mode $s$ the particle has been prepared for,
and the corresponding output $o_s$.
The input $i$ determines the ``context'' (a term borrowed from quantum logic)
of measurement,
whereas the output value $1\le o_s\le n$ defines the actual outcome.
${\rm RANDOM}(O)$ is a function whose value is a random element of $O$.

This straightforward implementation may not be regarded very elegant or even appropriate,
since it contains a random function in the definition of a finite automaton.
A conceptually more appealing Ansatz might be to get rid of the case $s\neq i$ in
which the automaton is prepared with a state information different from the one being retrieved.
One may speculate that the randomization of the output effectively originates from the
environment which ``translates'' the ``wrong'' question into the one which can be
answered by the automaton (which is constrained by $s=i$) at the price of randomization.
It also remains open if the automaton analogue to entanglement is merely an automaton
which cannot be decomposed into parallel single automata.

The formalism developed for quantum information in base $n$ defined by state partitions
can be fully applied to finite automata \cite{svozil-2002-statepart-prl}.
A $k$-particle system whose information is in base $n$
is described by $k$ nits which can be characterized by $k$
comeasurable partitions of the product state of the single-particle
states with $n$ elements each; every such element has $n^{k-1}$ elements.
Every complete set of comeasurable nits has the property that
(i) the set theoretic intersection of any $k$ elements of $k$ different
partitions is a single particle state, and (ii) the union of all these $n^k$
intersections is just the set of single particle states.
The set theoretic union of all elements of a complete set of comeasurable nits
form a state partition.
The set theoretic union of all of the $n^k!$ partitions
(generated by permutations of the product states)
form an automaton partition logic
corresponding to the quantum system.

\begin{figure}
\begin{center}
\tabcolsep 0 cm
\begin{tabular}{ccc}
 \includegraphics[width=4.1cm]{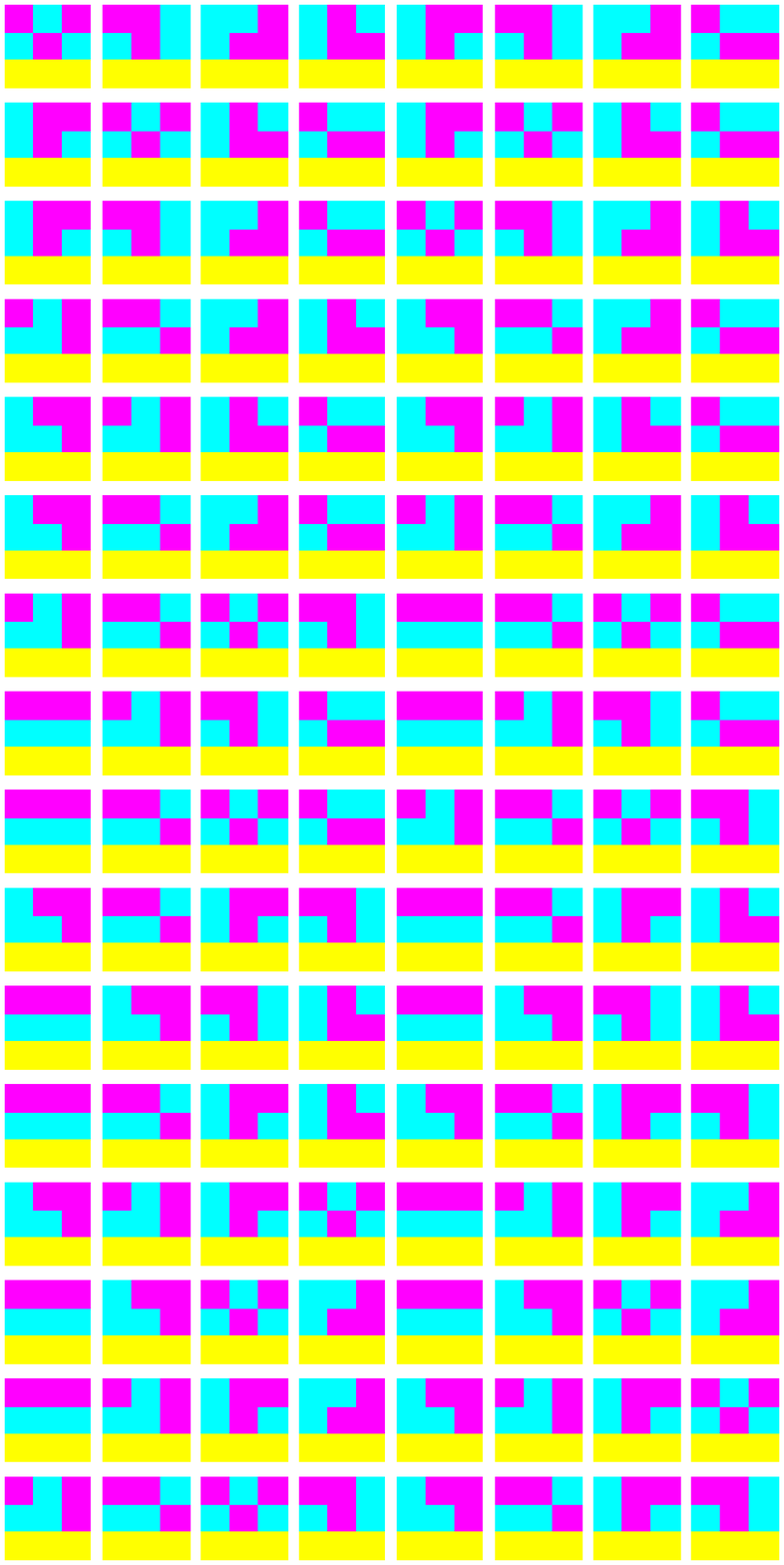} &
 \includegraphics[width=4.1cm]{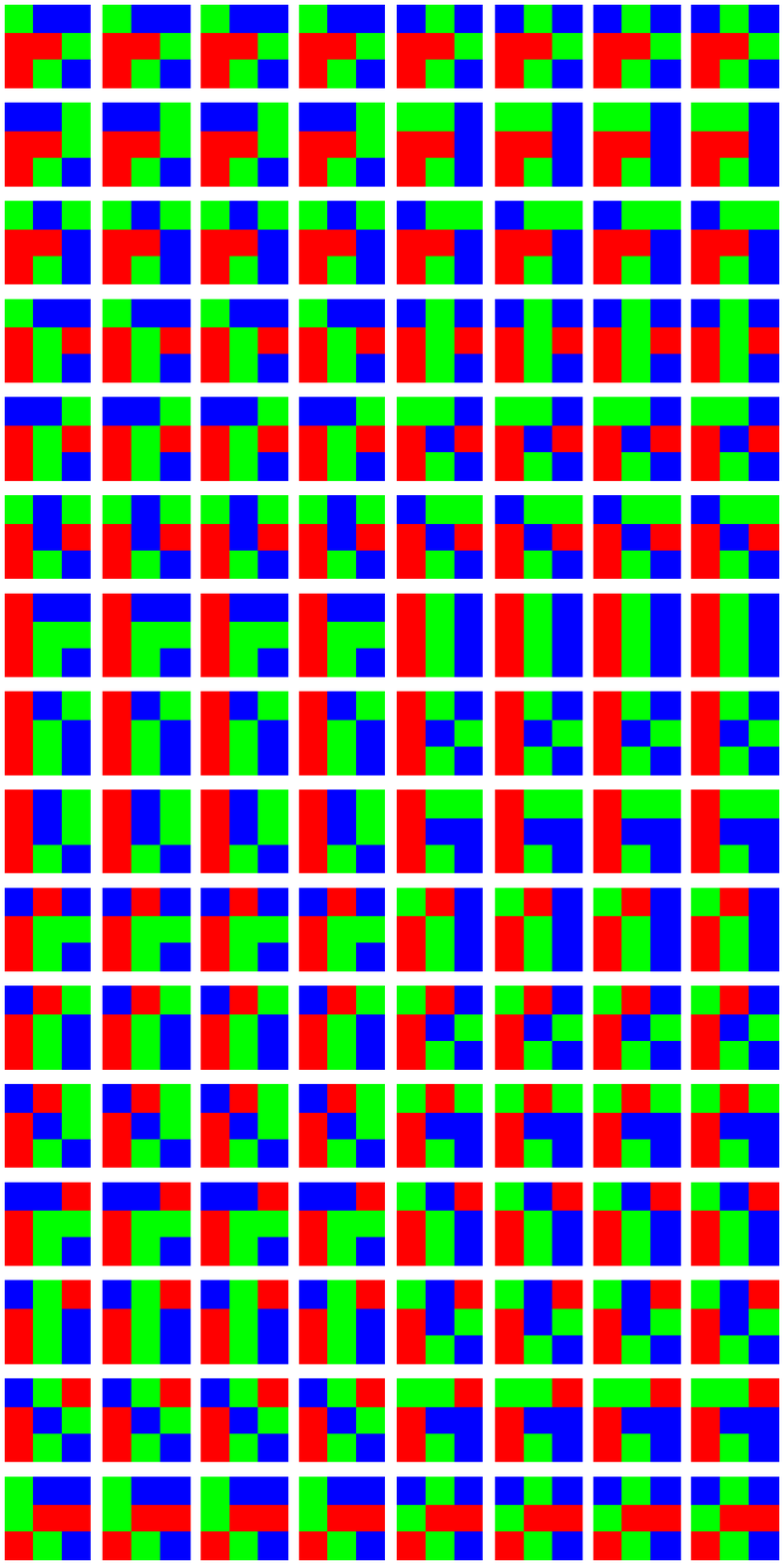} &
 \includegraphics[width=4.1cm]{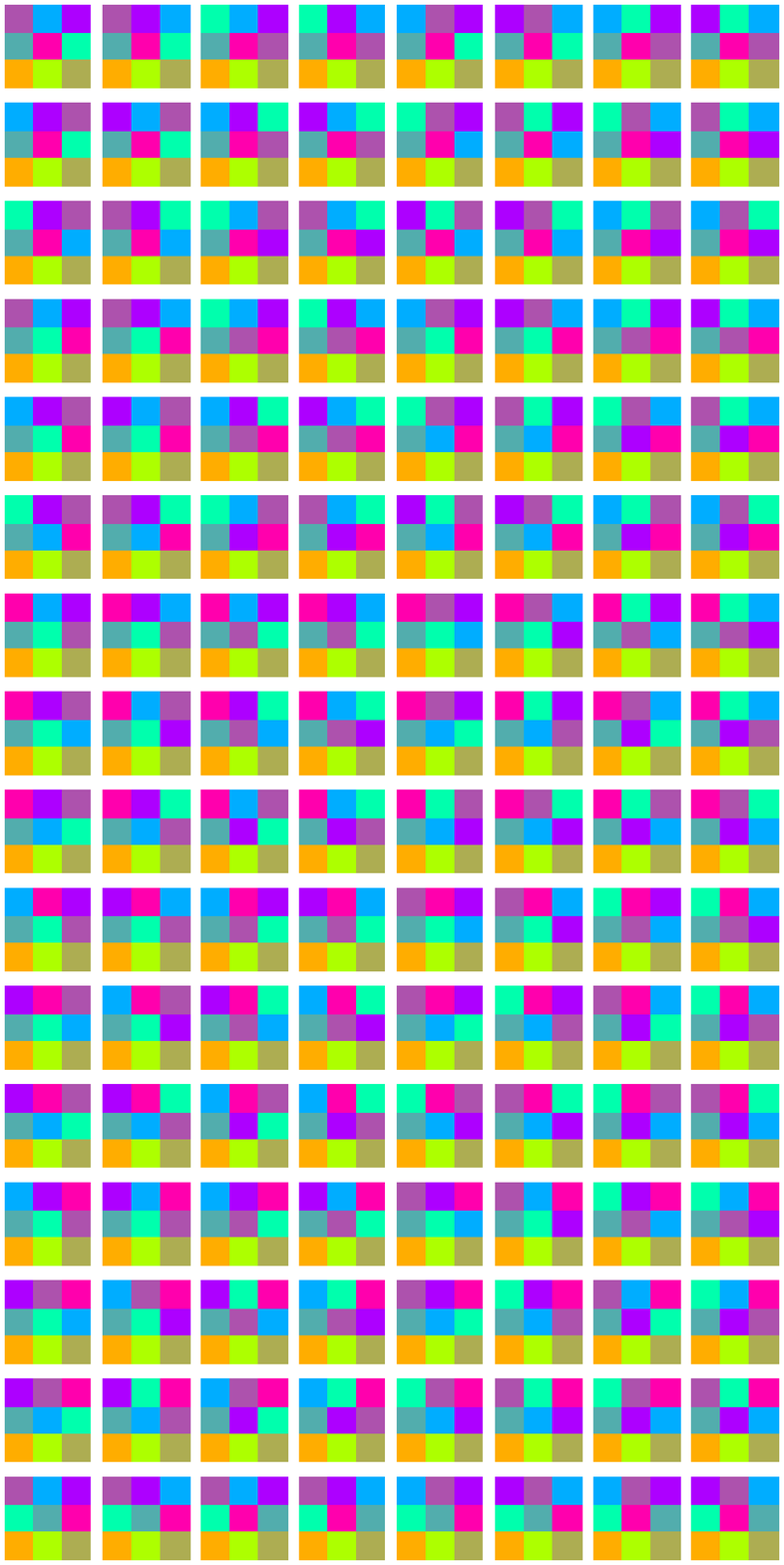}\\
trit 1&trit 2& trits 1\&2
\end{tabular}
\end{center}
\caption{Two trits yield a unique tessellation of the two-particle product state space; the first cases of Eqs.
(\ref{2002-kyoto-lb})---(\ref{2002-kyoto-lf}).\label{2002-kyoto-f1}}
\end{figure}

We shall demonstrate this construction with the case $k=2$ and $n=3$.
Suppose the product states are labeled from $1$ through $9$.
A typical element is formed by
$\{\{1,2,3\},\{4,5,6\},\{7,8,9\}\}$ for the first trit, and
$\{\{1,4,7\},\{2,5,8\},\{3,6,9\}\}$ for the second trit. Since those trits are comeasurable,
they can be united to form the first partition
$
\{
\{1,2,3\},\ldots ,\{3,6,9\}
\}
$.
The complete set of $9!/(2\cdot 3!\cdot 3!)= 5040$
different two-trit sets can be evaluated numerically; i.e.,
in lexicographic order,
%
%
%
\begin{eqnarray}
&\{\{\{1, 2, 3\}, \{1, 4, 5\}, \{2, 6, 7\}, \{3, 8, 9\}, \{4, 6, 8\}, \{5, 7, 9\}\},  &\label{2002-kyoto-lb}    \\
&\{\{1, 2,   3\}, \{1, 4, 5\}, \{2, 6, 7\}, \{3, 8, 9\}, \{4, 6, 9\}, \{5, 7, 8\}\},  &\\
&\vdots         \nonumber                                                             &             \\
&\{\{1,2,3\},\{\{1,4,7\},\{2,5,8\},\{3,6,9\},\{4,5,6\},\{7,8,9\}\} ,                  &     \\
&\vdots         \nonumber                                                             &              \\
&\{\{1, 6,   9\}, \{1, 7, 8\}, \{2, 4, 9\}, \{2, 5, 7\}, \{3, 4, 8\}, \{3, 5, 6\}\},  &    \\
&\{\{1, 6,   9\}, \{1, 7, 8\}, \{2, 4, 9\}, \{2, 5, 8\}, \{3, 4, 7\}, \{3, 5, 6\}\}\}.& \label{2002-kyoto-lf}
\end{eqnarray}
The associated partition logic is the horizontal sum of 5040
Boolean algebras with nine atoms (i.e., $2^9$) and corresponds to a rather
elaborate but structurally simple automaton with $9$ states $\{1,\ldots ,9\}$, $5040$ input symbols
$\{1,\ldots ,5040\}$ and $9$ output symbols $\{1,\ldots ,9\}$.
Every one of the Boolean algebras is the quantum analogue of a particular set of comeasurable
propositions associated with the two trits.
Together they form a complete set of trits for the two-particle three-state quantized case.
A graphical representation of the state single-particle state space tesselation
is depicted in Fig \ref{2002-kyoto-f1}.

\section{Applicability}

Despite some hopes, for instance stated by Einstein \cite[p.~163]{ein1},
to express finite, discrete physical systems by finite, algebraic theories,
a broader acceptance of automata models in physics would
require concrete, operationally testable consequences.
One prospect would be to search for phenomena which cannot happen according to
quantum mechanics but are realizable by finite automata.
The simplest case is characterized by a Greechie hyperdiagram of triangle form,
with three atoms per edge. Its automaton partition logic is given by
\begin{equation}
\{
\{\{1\},\{2\},\{3,4\}\},
\{\{1\},\{2,4\},\{3\}\},
\{\{1,4\},\{2\},\{3\}\}
\}.
\end{equation}
A corresponding Mealy automaton is
$\langle \{1,2,3,4\},\{1,2,3\},\{1,2,3\},\delta =1 ,\lambda \rangle$, where
$
\lambda (1,1)=
\lambda (3,2)=
\lambda (2,3)=
1
$,
$
\lambda (3,1)=
\lambda (2,2)=
\lambda (1,3)=
2
$, and
$
\lambda (2,1)= \lambda (4,1)=
\lambda (1,2)= \lambda (4,2)=
\lambda (3,3)= \lambda (4,3)=
3
$.

Another potential application is the investigation of the ``intrinsic
physical properties'' of virtual realities in general, and computer games in particular.
Complementarity and the other discussed features are robust
and occur in many different computational contexts, in particular
if one is interested in the intrinsic ``look and feel'' of computer animated worlds.


\end{document}